\documentclass[12pt]{iopart}

\usepackage{iopams}
\usepackage{graphicx}
\usepackage[usenames]{color}

\newcommand{\eqref}[1]{(\ref{#1})}
\newcommand{\be}{\begin{equation}}
\newcommand{\ee}{\end{equation}}
\newcommand{\bea}{\begin{eqnarray}}
\newcommand{\eea}{\end{eqnarray}}
\newcommand{\umg}{(1-\gamma)}

\begin{document}

\title{Velocity fluctuations in a one dimensional Inelastic Maxwell
model.}

\author{ G.Costantini$^1$, U. Marini Bettolo Marconi$^1$ and A. Puglisi$^2$ }
\address{$^1$ INFM-SOFT, 
Dipartimento di Fisica, Universit\`a di Camerino,
Via Madonna delle Carceri, 62032 , Camerino, Italy}
\address{$^2$ Dipartimento di Fisica, Universit\`a La Sapienza,                
Piazzale A. Moro 2, 00185 Roma, Italy}

\ead{giulio.costantini@unicam.it}

\begin{abstract}
We consider the velocity fluctuations of a system of particles
described by the Inelastic Maxwell Model.  The present work extends
the methods, previously employed to obtain the one-particle velocity
distribution function, to the study of the two particle correlations.
Results regarding both the homogeneous cooling process and the steady
state driven regime are presented.  In particular we obtain the form
of the pair correlation function in the scaling region of the
homogeneous cooling process and show that some of its moments
diverge. This fact has repercussions on the behavior of the energy
fluctuations of the model.
\end{abstract}
\maketitle

\section{Introduction}
\label{Introduction}
In recent years the understanding of the physics of granular materials
has taken great strides. The dynamical properties of particles
experiencing mutual inelastic collisions has been thoroughly studied
experimentally, theoretically and by computer simulation.  Such an
effort has lead to the discovery and to the formulation of new
phenomena and properties~\cite{introgas}.
Among these properties a special place is occupied by the homogeneous
cooling state (HCS), i.e. the state achieved by a granular gas,
initially in motion, under the effect of the energy loss caused by
inelastic collisions.  Loosely speaking the HCS plays for granular
gases a role analogous to the Maxwellian for molecular elastic
gases. Although the properties of the HCS are known in detail, its
explicit form can be obtained as series expansion only for some
specific models such as the inelastic hard-sphere model
(IHS)~\cite{brey97}.

The prototype model for the study of granular systems is represented
by an assembly of smooth inelastic hard spheres, characterized by a
constant coefficient of normal restitution.  For such a model various
authors have derived the Boltzmann and the Boltzmann-Enskog equations
describing the evolution of the reduced one-particle velocity
distribution~\cite{noije}.  However, since these approaches remain
mathematically hard to solve, a simpler mathematical model, the
Inelastic Maxwell Model (IMM), where the collision rate between the
particles is assumed to be independent of the relative velocity of the
colliding pair, has been put forward.  In this model the spatial
structure is neglected and only the velocity of the particles
specifies the state of the system.
The IMM is nevertheless useful and studied because it lends itself
to analytical solution in one dimension, thus providing a benchmark to test 
approximate treatments~\cite{Baldassarri}.
In the homogeneous free cooling case \cite{Pugliomixturesundriven}, 
the evolution equation for the
velocity distribution has a scaling solution that can be expressed in
an analytical closed form, with high energy tails described by an
algebraic decay: the exponent does not depend on the restitution
coefficient. Moments of the velocity distribution exhibit multiscaling
asymptotic behavior~\cite{BenNaim}.  The Inelastic Maxwell Model is
quite simplified with respect to inelastic hard spheres and other
realistic models of dilute granular materials, nevertheless in the
past it has been considered an important starting point for granular
kinetic theories~\cite{Santos}. As stated by Ernst and Brito~\cite{Brito}:
``What
harmonic oscillators are for quantum mechanics, and dumb-bells for
polymer physics, is what elastic and inelastic Maxwell models are for
kinetic theory.''


Most of the literature on the kinetic theory of granular gases focuses
on the single particle distribution function. This is in analogy with
the relevance that the Molecular Chaos approximation has for molecular
(i.e. elastic) gases. On the other hand, the inelasticity of
collisions in granular gases makes this assumption more delicate:
numerical and experimental evidences show a stronger tendency of
granular systems to enhance correlations, often appearing in the form
of spatial structures~\cite{Goldhirsch,Sela,Andreabaldassarri}.
Fluctuations have been investigated by various authors 
\cite{vanNoije,Brey98,Soto}.
However,
only recently the two-particles distribution function has come under
scrutiny, in particular by Brey and coworkers who considered its
application to the study of the energy fluctuation in the homogeneous
cooling state of inelastic hard spheres~\cite{Brey}. Their study focuses on
the effect of inelasticity  on the $1/N$ deviations
from Molecular Chaos.  Here, our aim is to apply similar analysis
to the one dimensional Inelastic Maxwell Model. This is interesting
because, with a few controlled approximations, one obtains the
asymptotic pair correlation function in a closed form, and all time
dependencies of its two-particles velocity moments, getting further
than the original work of Brey et al., where only the asymptotic
moments, in particular those required to calculate energy
fluctuations, were explicitly obtained.

This paper is organized as follows. In section \ref{Evolution} 
the evolution
equation for the IMM is presented and the equations for the various
distribution functions introduced.  In section \ref{Power}  the dynamical
equations are solved for the moments of the single and two-particles
distribution functions. The asymptotic scaling state is discussed, for
the pair correlation function, in section \ref{Scaling} . 
Finally, in Section \ref{Driven}  the
effects of an external driving is considered and 
in~\ref{Conclusion} the concluding remarks are presented.

\section{Evolution equations for the distribution functions}
\label{Evolution}
We consider a system of N particles, each 
characterized by a scalar velocity $v_i$,
with $i=1,...,N$. The Inelastic Maxwell Model assumes that the
state $\Gamma=(v_1,v_2,..,v_N)$ is modified by elementary collision events, 
realized by changing the velocities $(v_i,v_j)$ of a 
randomly selected  pair of particles  according to the rule:
\begin{eqnarray}
v_i'&=&\gamma v_i+(1-\gamma)v_j 
\label{col2} \\ \nonumber v_j'&=& (1-\gamma)v_i+\gamma v_j  \qquad 
\end{eqnarray}
where $\gamma=\frac{1-\alpha}{2}$ and $\alpha$ is the coefficient 
of restitution.
The system cools down because in each 
collision an amount $\Delta E$ of kinetic energy, given by
\begin{equation}
\Delta E=-\frac{m}{4}(1-\alpha^2)(v_i'-v_j')^2, \qquad
\label{u1}
\end{equation}
is dissipated,  where $m$ is the mass of a particle.
Since the IMM is not endowed with
a spatial structure such a cooling process is homogeneous. 
An observable $A(\Gamma(t))$ evolves according to 
\begin{equation}
A(\Gamma(t))=\exp(t {\cal L}A(\Gamma(0)))
\label{u2}
\end{equation} 
where the generator ${\cal L}$ is 
\begin{equation}
{\cal L}=\frac{1}{2}\sum_{i=1}^{N}\sum_{j\neq i}^{N} T(v_i,v_j)
\end{equation}
and the operator $T$ acts on an arbitrary function $S(v_i,v_j)$
of the velocities of particles $i$ and $j$ in the following way:
\begin{equation}
T(v_i,v_j)S(v_i,v_j)=S(v_i',v_j')-S(v_i,v_j)
\end{equation}
The ``time'' $t$ is a collision counter and represents the clock
of the model.

Following closely the derivation presented by Brey et al.\cite{Brey},
in order to set up the evolution equations for the system, we introduce 
the following distribution functions
\begin{equation}
F_1(u_1,t)=
\sum_{i=1}^{N}\langle \delta(u_1-v_i(t))\rangle
\label{f1}
\end{equation} 
\begin{equation}
F_2(u_1,u_2,t)=
\sum_{i=1}^{N}\sum_{j\neq i}^{N}\langle \delta(u_1-v_i(t))
\delta(u_2-v_j(t))\rangle
\label{f2}
\end{equation}
\begin{equation} 
F_3(u_1,u_2,u_3,t)=
\sum_{i=1}^{N}\sum_{j\neq i}^{N}\sum_{k\neq i,j}^{N}
\langle \delta(u_1-v_i(t))
\delta(u_2-v_j(t))\delta(u_3-v_k(t))\rangle
\label{f3}
\end{equation} 
where $\langle \cdot \rangle$ stands for an average over an
ensemble of trajectories with different initial conditions (in section
\ref{Driven}, where the effect of a thermal bath will be considered, this will
be an average over realizations of the noise).

A hierarchy of equations can be derived for these distribution functions, 
whose first two equations read:
\begin{equation}
\frac{d}{dt}F_1(u_1,t)=\int du_2 {\overline T}(u_1,u_2)F_2(u_1,u_2,t)
\label{e1}
\end{equation} 
and
\begin{eqnarray}
\frac{d}{dt}F_2(u_1,u_2,t)&=& {\overline T}(u_1,u_2)F_2(u_1,u_2,t)+\nonumber \\
&+&\int du_3 [{\overline T}(u_1,u_3)+{\overline T}(u_2,u_3)]F_3(u_1,u_2,u_3,t)
\label{e2}
\end{eqnarray} 
where the inverse binary collision operator, ${\overline T}$, is defined 
for a generic function $S(u_i,u_j)$ of the velocities by the rule:
\begin{equation}
{\overline T}(u_i,u_j)S(u_i,u_j)=\frac{1}{\alpha}
S(u_i^{*},u_j^{*})-S(u_i,u_j)
\end{equation}
which transforms the velocities  $(u_i,u_j)$ into their
pre-collisional values $u_i^{*},u_j^{*}$,
obtained by inverting eq.~\eqref{col2}.
Following a standard statistical procedure we consider the following
decompositions of the distribution functions:
\begin{equation}
F_2(u_1,u_2,t)=F_1(u_1,t)F_1(u_2,t)+G_2(u_1,u_2,t)
\label{cl1}
\end{equation}
\bea
F_3(u_1,u_2,u_3,t)&=&F_1(u_1,t)F_1(u_2,t)F_1(u_3,t)+ G_2(u_1,u_2,t)F_1(u_3,t)
\nonumber\\
&+&G_2(u_1,u_3,t)F_1(u_2,t)+G_2(u_2,u_3,t)F_1(u_1,t)\nonumber\\&+&
G_3(u_1,u_2,u_3,t)
\label{cl2}
\eea
After substituting these expressions into \eqref{e1} and \eqref{e2}
and dropping the term containing $G_3$ we obtain a pair of closed equations
for $F_1$ and $G_2$.
Before proceeding further  we also define, for later convenience, the following
normalized distributions:
\begin{equation}
f_1(u_1,t)=\frac{1}{N}F_1(u_1,t)
\label{bf1}
\end{equation} 
and 
\begin{equation}
h_2(u_1,u_2,t)=
\frac{1}{N}G_2(u_1,u_2,t)
\label{bf3}
\end{equation}
which obey the sum rules 
\bea
&&\int  du_1 f_1(u_1,t)=1 \\
&&\int\int  du_1 du_2 h_2(u_1,u_2,t)=-1\\
&&\int du_2 h_2(u_1,u_2,t)=- f_1(u_1 ,t)
\label{sumrules}
\eea
Hence eqs. \eqref{e1} and \eqref{e2} can be rewritten as
\begin{equation}
\frac{d}{d\tau}f_1(u_1,\tau)=\int du_2 {\overline T}(u_1,u_2)
[f_1(u_1,\tau)f_1(u_2,\tau)+\frac{1}{N}h_2(u_1,u_2,\tau)]
\label{ce1}
\end{equation} 
and
\bea
&\frac{d}{d\tau}&h_2(u_1,u_2,\tau)= {\overline T}(u_1,u_2)
[f_1(u_1,t)f_1(u_2,\tau)+ \frac{1}{N}h_2(u_1,u_2,\tau)]\nonumber\\
&+&\int du_3 {\overline T}(u_1,u_3)[h_2(u_1,u_2,\tau)f_1(u_3,\tau) +
h_2(u_2,u_3,t)f_1(u_1,\tau)]\nonumber\\
&+&\int du_3 {\overline T}(u_2,u_3)]
[h_2(u_1,u_2,\tau)f_1(u_3,\tau) +
h_2(u_1,u_3,\tau)f_1(u_2,\tau)]
\label{ce2}
\eea 
where we have redefined the time variable $\tau=N t$.

In order to solve \eqref{ce1} and \eqref{ce2} we
slightly generalize the method, originally introduced 
by Bobylev \cite{Bobylev} and consider the following Fourier 
transforms of the distributions $f_1$ and $h_2$
\bea
&&\hat f_1(k_1,t)=\int  du_1 e^{i k_1 u_1} f_1(u_1,t) \\
&&\hat h_2(k_1,k_2,t)=\int \int  
du_1 du_2  e^{i k_1 u_1+i k_2 u_2} h_2(u_1,u_2,t)
\label{bobylev}
\eea
The function $\hat h_2(k_1,k_2,t)$ is symmetric and
has the property $\hat h_2(k_1,0,t)=-f_1(k_1,t)$ as a consequence of the
sum rule  \eqref{sumrules}.
Substituting these expressions into \eqref{ce1} and \eqref{ce2} we find:
\bea
\frac{d}{d\tau}\hat f_1(k_1,\tau)&=&
\hat f_1(\gamma k_1,\tau)\hat f_1(\umg k_1,\tau)
-\hat f_1(k_1,\tau)\hat f_1(0,\tau)\nonumber\\
&+&\frac{1}{N}[\hat h_2(\gamma k_1,\umg k_1,\tau)-\hat h_2(k_1,0,\tau)]
\label{de1}
\eea 
and
\bea
&\frac{d}{d\tau}&\hat h_2(k_1,k_2,\tau)=
\frac{1}{N}[\hat h_2(\gamma k_1+\umg k_2,\gamma k_2+\umg k_1,\tau)-
\hat h_2(k_1,k_2,\tau)]\nonumber\\
&+&\hat f_1(\gamma k_1+\umg k_2,\tau)\hat f_1(\gamma k_2+\umg k_1,\tau)
-\hat f_1(k_1,\tau)\hat f_1(k_2,\tau)\nonumber\\
&+&(1+P_{12})[\hat f_1(\umg k_1,\tau)\hat h_2(\gamma k_1,k_2,\tau)-
\hat f_1(0,\tau)\hat h_2(k_1,k_2,\tau)\nonumber\\
&+&\hat f_1(\gamma k_1,\tau)\hat h_2(k_2,\umg k_1,\tau)-
\hat f_1(k_1,\tau)\hat h_2(k_2,0,\tau)]\nonumber\\
\label{de2}
\eea 
where $P_{12}$ exchanges the index $1$ and $2$. 


It is possible to connect some elements of the pair distribution
function to observable properties. To this purpose we
consider the distribution functions, $\Pi_d(V)$ and $\Pi_s(W)$,
of the difference of the velocities $V=(u_1-u_2)$, and of the
sum  $W=(u_1+u_2)$, which are obtained by marginalizing the
distribution $f_2(u_1,u_2,t)=f_1(u_1,t)f_1(u_2,t)+h_2(u_1,u_2,t)/N$ 
according to the transformations:
\be
\Pi_d(V)=\int\int du_1 d u_2 f_2(u_1,u_2)\delta (V-(u_1-u_2))
\label{distrdiff}
\ee
and 
\be
\Pi_s(W)=\int\int du_1 d u_2 f_2(u_1,u_2)\delta (W-(u_1+u_2))
\label{distrsum}
\ee
We take, now, Fourier-Bobylev transforms of both distribution functions
and find
\be
\hat\Pi_d(k,\tau)=\hat f_1(k,\tau)\hat f_1(-k,\tau)
+\frac{1}{N}\hat h_2(k,-k,\tau)
\ee
Similarly
\be
\hat\Pi_s(k,\tau)= \hat f_1(k,\tau)\hat f_1(k,\tau)+\frac{1}{N}
\hat h_2(k,k,\tau)
\ee
Of course, the correction is of order $1/N$ and vanishes for infinite
systems.

In the following we shall assume $1/N << 1$ and drop the corresponding
terms in \eqref{de1} and \eqref{de2} (see Appendix for a discussion of
this approximation). Hence, eq. \eqref{de1} reduces to the standard
equation of the one-dimensional IMM \cite{BenNaim} and decouples from
the evolution equation for $\hat h_2$.


\section{Power series solution.}
\label{Power}
The distribution functions can be expanded into their moments as follows:
\bea
&&M_n(\tau)= \int  du_1 u_1^n f_1(u_1,\tau) \\
&&Q_{mn}(\tau)=\int\int  du_1 du_2 u_1^m u_2^n h_2(u_1,u_2,\tau),
\label{moments}
\eea
obtaining:
\bea
&&\hat f_1(k_1,\tau)=\sum_{n=0}^{\infty}\frac{(i k_1)^n}{n!}M_n(\tau) \\
&&\hat h_2(k_1,k_2,\tau)=\sum_{m,n=0}^{\infty}
\frac{(i k_1)^m}{m!}\frac{(i k_2)^n}{n!}Q_{mn}(\tau). \nonumber\\
\label{bobylev2}
\eea
Inserting these expansions into eqs.~\eqref{de1} and \eqref{de2} we 
first recover
the moments, $M_i$, evaluated by Ben-Naim
and Krapivski~\cite{BenNaim} and given by:
\bea
&& M_0(\tau)= 1\\
&&M_1(\tau)= 0\\
&&M_2(\tau)=M_2(0)e^{-a_2\tau} \\
&&M_3(\tau)=M_3(0)e^{-a_3\tau}  \\
&&M_4(\tau)=[M_4(0)+3 M^2_2(0)]e^{-a_4\tau}-3 M^2_2(\tau) 
\label{fmoments}
\eea with the coefficients given by $a_n=1-\umg^n-\gamma^n$,
$a_{24}=6\gamma^2\umg^2$ and $\zeta=\gamma(1-\gamma)$.  Notice also
that $a_2=2\zeta$ and $a_3=3 \zeta$, and $a_4-2 a_2= -2 \zeta^2$.  In
addition, we obtain the moments of $h_2$ using the conditions that the
initial velocities are indipendently distributed (with a constraint on
the total momentum $M_1(\tau)=0$) and the one-particle distribution is
even: \bea Q_{n0}(\tau)&=& Q_{0n}(\tau)= -M_n(\tau)\\ Q_{n1} (\tau)&=&
Q_{1n} (\tau)=-M_{n+1}(\tau)\\
Q_{22}(\tau)&=&[Q_{22}(0)-M_4(0)]e^{-2a_2\tau}+M_4(\tau)
\label{qmoments}
\eea

We can, now, compute the energy fluctuations since
\bea
\langle E^2(\tau)\rangle-\langle E(\tau)\rangle^2&=&\frac{Nm^2}{4}\Bigl\{
\int \int  du_1 du_2 u_1^2 u_2^2 h_2(u_1,u_2,\tau)+\int  du_1 u_1^4 
f_1(u_1,\tau)\Bigl\}\nonumber\\
&=&\frac{Nm^2}{4}
\Bigl\{ Q_{22}(\tau)+M_{4}(\tau)
\Bigl\},
\eea
having defined the total energy as $E(\tau)=\frac{m}{2}\sum_{i=1}^N v_i^2(\tau)$.
Recalling the kinetic definition of granular temperature 
$T_g(\tau)=mM_2(\tau)=2\langle E(\tau) \rangle/N$, one has that
\begin{equation}
\frac{\langle E^2(\tau)\rangle-\langle 
E(\tau)\rangle^2}{T_g(\tau)^2}=\frac{N}{4}[A+B\exp(2\zeta^2 \tau)]
\end{equation}
with $A=\frac{Q_{22}(0)- M_4(0)}{M^2_2(0)}-6$ and $B=2 \frac{M_4(0)+3
M^2_2(0)}{ M^2_2(0)}$.  Therefore the energy fluctuations decay at a
slower rate than the square of the average energy. The situation is
analogous to what happens to the fourth moment of the distribution
function, which also diverges if rescaled by the square of the second
moment.  On the other hand, we notice that the energy fluctuations,
scale proportionally to $N$, as in non critical systems: this means
that a thermal capacity can always be defined, but it grows with
time. This is different from what happens in the homogeneous cooling
of inelastic hard spheres, as discussed by Brey et al.\cite{Brey},
where the ratio between energy fluctuations and the square of average
energy is constant in the HCS scaling state.

\section{Fluctuations around the scaling solution.}
\label{Scaling}
It is well known that eq. \eqref{de1} for large N, possesses a scaling
solution\cite{Baldassarri}, where the only time dependence occurs via
the combination $q_1(\tau)=k_1 v_0(\tau)$, i.e.
$\hat{f}_1(k_1,\tau)=\xi_0(q_1)$, with $v_0(\tau)=\sqrt{M_2(\tau)}$
the thermal velocity. Using such a variable the evolution equation for
the distribution function takes the scaling form
\begin{equation}
-\zeta q_1\frac{d}{d q_1}\xi_0(q_1)=\xi_0(\gamma q_1)\xi_0(\umg q_1)
-\xi_0(q_1)\xi_0(0)
\label{sc2}
\end{equation}
which has the solution
\begin{equation}
\xi_0(q_1)= (1+|q_1|)e^{-|q_1|}.
\label{sc1}
\end{equation}
Its small $q_1$ singularity reflects the inverse power law tails of
the corresponding velocity distribution function~\cite{Brito},
$\phi_0(c)$ which is obtained by applying the inverse Bobylev-Fourier
transform to eq. \eqref{sc1} with the result: 
\be
\phi_0(c)=\frac{2}{\pi}\frac{1}{[1+c^2]^2}, 
\ee 
with $c=u/v_0(t)$. Note that  the complete time-dependent velocity
distribution reads $f_1(u_1,t)=\phi_0[u_1/v_0(t)]/v_0(t)$.  We wish to
consider, now, the fluctuations around the scaling solution.
We first define the functions $\xi_i$ with $i=1,2$ defined as
\be
\xi_1(q_1)=(|q_1|+q_1^2)e^{-|q_1|},\qquad
\xi_2(q_1)=q_1^2 e^{-|q_1|}
\ee
and the linearized Maxwell-Boltzmann
operator $\Lambda_1\equiv\Lambda(q_1;\xi_0(q_1))$ as:
\bea
\Lambda_1\psi(q_1) &=&\zeta q_1\frac{d}{d q_1}\psi(q_1)
+\xi_0(\umg q_1)\psi_2(\gamma q_1)-\xi_0(0)\psi(q_1)\nonumber\\
&+&\xi_0(\gamma q_1)\psi(\umg q_1)-\xi_0(q_1)\psi(0).
\label{blin1}
\eea One can see that \bea \Lambda_1\xi_0=\zeta\xi_2,\qquad
\Lambda_1\xi_1=\zeta\xi_1,\qquad \Lambda_1\xi_2=0 \eea and conclude
that $\xi_1$ and $\xi_2$ are eigenfunctions of $\Lambda_1$
corresponding to the eigenvalues $\zeta$ and $0$ respectively, whereas
$\xi_0$ is not eigenfunction of $\Lambda_1$. Interestingly, the
Bobylev-Fourier transforms of $\xi_2$ and $\xi_1$ read, respectively:
\bea \phi_2(c)&=&\frac{d}{dc}(c\phi_0(c))\nonumber\\
\phi_1(c)&=&\frac{d}{dc}\frac{1}{\pi}\frac{c}{1+c^2}+\phi_2(c).  \eea
Remarkably, these two eigenfunctions have a similar structure to that
of two of the three eigenfunctions found by Brey and co-workers, see
Equations (65) in their paper~\cite{Brey}.  This
similarity is however incomplete, since in~\cite{Brey} the
eigenfunctions of the linearized Boltzmann operator were identified to
be the hydrodynamic modes. Up to our knowledge, a study of
hydrodynamic spectrum for the Inelastic Maxwell Model is
missing (it has been performed, for Inelastic Hard Spheres
in~\cite{brey2}) and therefore we are not able to make a similar
connection, neither to find the third eigenfunction necessary for completing
the analogy.

In order to determine the pair correlation function 
we, now, rewrite eq.\eqref{de2} in the scaling form
\be
[\Lambda_1 +\Lambda_2]\chi_2(q_1,q_2)
=U(q_1,q_2)
\label{qre3}
\ee
with
$\chi_2(q_1,q_2)\equiv\hat h_2(k_1,k_2,\tau)$,
\be
U(q_1,q_2)
=-\xi_0(\gamma q_1 +\umg q_2)\xi_0(\umg q_1+\gamma q_2)
+\xi_0(q_1)\xi_0(q_2)
\label{qre4}
\ee
and having defined the operator $\Lambda_2$ as identical to $\Lambda_1$ but acting upon the variable $q_2$.
If $q_1 q_2\geq0$ formula \eqref{qre4} can be cast in the form
\bea
U(q_1,q_2) 
&=&-\zeta[\xi_2(q_1)\xi_0(q_2)+\xi_0(q_1)\xi_2(q_2)\\
&+&\xi_1(q_1)\xi_2(q_2)+\xi_2(q_1)\xi_1(q_2)
-2\xi_1(q_1)\xi_1(q_2)] \nonumber
\label{qre6}
\eea
so that we easily find a solution 
\bea
\chi_2(q_1,q_2)&=&\theta(q_1q_2)\bigl(-\xi_0(q_1)\xi_0(q_2)+\xi_1(q_1)\xi_1(q_2)
+\tilde{C}_{22}\xi_2(q_1)\xi_2(q_2)\nonumber\\
&-&[\xi_1(q_1)\xi_2(q_2)+\xi_1(q_2)\xi_2(q_1)]\bigl)
\label{qcoef2}
\eea
where $\theta(x)$ is the Heaviside step function and $\tilde{C}_{22}$ is an arbitrary constant which may  be fixed using the boundary conditions.
For $q_1 q_2<0$ we could not find an exact solution. The
most reasonable path to reach an expression for $\chi_2(q_1,q_2)$ in
this part of the $q_1,q_2$ plane is therefore to expand the r.h.s. of
Eq.~\eqref{qre3} on the basis $\xi_i$ (see Appendix \ref{appendix1}
for details) and assume \be \chi_2(q_1,q_2)=\theta(-q_1q_2)
\sum_{m,n}^2 C_{mn}\xi_m(q_1)\xi_n(q_2).
\label{chi2neg}
\ee This procedure,  which is justified only as an
approximate ``continuation'' of the solution in the $q_1 q_2>0$
region, and will be checked for consistency at the end of this
section, leads to the following equation for the coefficients
$C_{mn}$: \be [\Lambda_1 +\Lambda_2]\sum_{m,n}^2
C_{mn}\xi_m(q_1)\xi_n(q_2) =\sum_{m,n}^2 T_{mn}\xi_m(q_1)\xi_n(q_2)
\label{ex1}
\ee
whose
approximate solution can be found by
expanding $T_{mn}$ in powers of $\zeta$ up to order $\zeta^3$ 
(i.e. for small inelasticity). The final result reads: 
\be
C_{00}=-1+\frac{3}{2}\zeta+45\zeta^2
\label{c00}
\ee
\be
C_{01}=C_{10}=-\frac{3}{2}\zeta-33\zeta^2
\label{c01}
\ee
\be
C_{11}=- 1+\frac{21}{2}\zeta+\frac{135}{2}\zeta^2
\label{c11}
\ee
\be
C_{12}=C_{21}=1-\frac{61}{2}\zeta-128\zeta^2
\label{c12}
\ee
\be
C_{02}=C_{20} = \frac{39}{2}\zeta+\frac{147}{2}\zeta^2
\label{c02}
\ee

Also in this case $C_{22}$ results arbitrary. 
Notice that the form of the pair correlation function $\chi_2(q_1,q_2)$
in the region
$q_1q_2<0$ now depends on the inelasticity through $\zeta$ and 
therefore is not universal as the single-particle distribution.

In the previous section we have seen that the rescaled energy fluctuations,
related to the moment $Q_{22}$,
diverge as $\tau\to\infty$. The situation is similar to that encountered
in the study of the single particle distribution function, where
the exponential increase of 
the rescaled fourth moment $M_4(t)/M_2(t)^2$ was the signature
of the fact that the fourth moment
of $\phi_0(c)$ diverges. Is the behavior of $Q_{22}$ 
the fingerprint of a similar behavior of $\chi_2(q_1,q_2)$? 
Indeed, the size of the energy fluctuations is controlled by
the small $(q_1,q_2)$ singularities of $\chi_2(q_1,q_2)$,
because these determine the high velocities tails of the correlations
\cite{Brito,Pugliomixturesundriven}.
In order to expose the presence of the tails of  $\chi_2$,
we isolate the most singular contribution to it, namely
the term proportional to $\xi_1(q_1)\xi_1(q_2)$ in eqs. 
\eqref{qcoef2} and \eqref{chi2neg}
\begin{eqnarray}
\chi_2^{sing}(q_1,q_2)&=&[\theta(q_1q_2)+\theta(-q_1q_2)C_{11}]
\xi_1(q_1)\xi_1(q_2)\nonumber \\
&\simeq& [\theta(q_1q_2)+\theta(-q_1q_2)C_{11}]|q_1||q_2|e^{-|q_1|-|q_2|},
\end{eqnarray}
After Fourier transforming to velocity space 
we realize that the pair correlation
function for large value of its arguments decays as:
\be
FT[\chi_2^{sing}]\simeq \frac{1}{c_1^2 c_2^2} 
\ee
so that the moment $c_1^2 c_2^2$ diverges. Such a result is the counterpart
of the divergence of the fourth moment of the single particle 
distribution function.

\begin{figure}[htbp]
\includegraphics[width=10cm,clip=true]{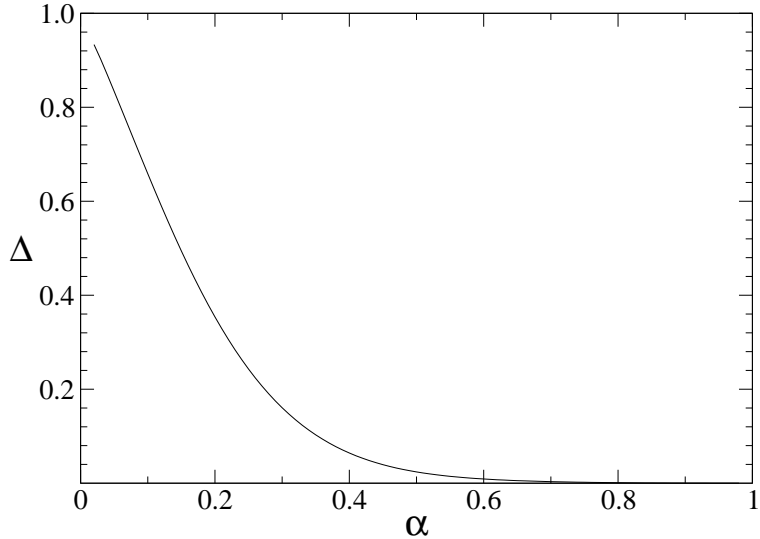}
\caption{The error $\Delta$ of eq. (\ref{errore}) as a function of $\alpha$.
\label{figerrore}}
\end{figure}

Before closing the present section we wish to comment the fact that
the projection introduces an error due to the truncation of the
expansion of $\tilde U(q_1,q_2)$ (see eq.\eqref{a7} ). 
How reliable is such an approximation? In order to check the error
we computed,  for various values of $\alpha$, the following quantity:
\be
\Delta=\frac{\int\int dq_1 dq_2 \Bigl[\tilde U(q_1,q_2)-\sum_{m,n=0}^2 
T_{mn}\xi_m(q_1)\xi_n(q_2)\Bigl]^2}
{\int\int dq_1 dq_2 \Bigl[\tilde U(q_1,q_2)\Bigl]^2}
\label{errore}
\ee which represents a measure of the relative error.  As we see from
fig. \ref{figerrore} the approximation becomes poorer and poorer as
$\alpha\to 0$. However, for not too low inelasticities the approximation is
reasonable  and in our opinion this justifies the above
procedure, in particular Eq.~\eqref{chi2neg}, for value of $\alpha \gtrsim 0.5$.

\section{Driven system}
\label{Driven}
Now, let us consider a system driven by an external Langevin heat-bath 
which has been considered by several authors
\cite{Pagnani,Andrea2002,Brito2002,Santos2003}.
The velocities of the particles, 
evolve between collisions according to an Ornstein-Uhlenbeck process:
\be
\frac{d v_i}{d t}=-\Gamma v_i(t)+\xi_i(t)
\ee
with 
\be
\langle\xi_i(t)\rangle=0, \qquad \langle\xi_i(t)\xi_j(t')\rangle=
2D\delta_{ij}\delta(t-t')
\ee 
The resulting equations for the distribution functions are:
\bea
\frac{d}{d\tau}\hat f_1(k_1,\tau)&=&-(D k_1^2+\Gamma k_1 \frac{d}{d k_1})
\hat f_1(k_1,\tau) \nonumber \\
&+&\frac{1}{\tau_c}\{\hat f_1(\gamma k_1,\tau)\hat f_1(\umg k_1,\tau) 
-\hat f_1(k_1,\tau)\hat f_1(0,\tau)\nonumber \\
&+&\frac{1}{N}[\hat h_2(\gamma k_1,\umg k_1,\tau)-\hat h_2(k_1,0,\tau)]\}
\label{bde1}
\eea 
where an arbitrary mean free $\tau_c$ time has now been introduced 
for dimensional reasons, and
\bea
\frac{d}{d\tau}\hat h_2(k_1,k_2,\tau)&=&
- [D(k_1^2+k_2^2)+\Gamma (k_1\frac{d}{d k_1}+k_2\frac{d}{d k_2})]
\hat h_2(k_1,k_2\tau)\nonumber\\
&+&\frac{1}{\tau_c}\{
\frac{1}{N}[\hat h_2(\gamma k_1+\umg k_2,\gamma k_2+\umg k_1,\tau)-
\hat h_2(k_1,k_2,\tau)]\nonumber\\
&+&\hat f_1(\gamma k_1+\umg k_2,\tau)\hat f_1(\gamma k_2+\umg k_1,\tau)
-\hat f_1(k_1,\tau)\hat f_1(k_2,\tau)\nonumber\\
&+&(1+P_{12})[\hat f_1(\umg k_1,\tau)\hat h_2(\gamma k_1,k_2,\tau)-
\hat f_1(0,\tau)\hat h_2(k_1,k_2,\tau)\nonumber\\
&+&\hat f_1(\gamma k_1,\tau)\hat h_2(k_2,\umg k_1,\tau)-
\hat f_1(k_1,\tau)\hat h_2(k_2,0,\tau)]\}\nonumber\\
\label{bde2}
\eea 
We look for steady state solutions, setting the time derivatives 
to zero.
By slightly modifying the method  employed to derive the moments
in the cooling case we obtain the moments in the non equilibrium steady
state regime. We find:

\begin{eqnarray}
M_{0}&=&1  \\
M_{2}&=&\frac{D\tau_c} {\Gamma\tau_c+\gamma(1-\gamma)}  \label{gt} \\
M_{4}&=&\frac{12 D\tau_c M_2+6 \gamma^2(1-\gamma)^2
M_{2}^2}{4\Gamma\tau_c+1-\gamma^4-(1-\gamma)^4} \nonumber \\ &=&\frac{12[\Gamma \tau_c+\gamma(1-\gamma)]+6\gamma^2(1-\gamma)^2}{4\Gamma\tau_c+1-\gamma^4-(1-\gamma)^4}M_2^2
\end{eqnarray}
 The granular temperature $T_g=mM_2$ is obtained from
equation~\eqref{gt}, yielding in the elastic case ($\gamma=0$)
$T_g=mD/\Gamma$ as expected. Furthermore, the last equations in the
elastic case becomes $M_4=3M_2^2$.

Similarly we find the coefficients of the moments of the pair
correlation $h_2$ using in eq. \eqref{bde2} the
expansion~\eqref{moments}
\begin{eqnarray}
Q_{n0}&=&Q_{0n}=-M_n  \\
Q_{11}&=&\frac{\gamma(1-\gamma)} {\Gamma\tau_c}M_2   \\
Q_{12}&=& Q_{21}=0\\
Q_{13}&=&Q_{31}=\frac{\gamma (1-\gamma)[1-2\gamma(1-\gamma)](3M_2^2+M_4)+
3\Gamma \tau_c Q_{11}^2+
6D\tau_c Q_{11}}{4\Gamma\tau_c+3\gamma(1-\gamma)}   \\ 
Q_{22}&=&\frac{\gamma^2(1-\gamma)^2(3M_2^2+M_4)-2D\tau_c M_2} 
{2\Gamma\tau_c+2\gamma(1-\gamma)} \nonumber \\&=&\frac{\gamma^2(1-\gamma)^2(C+3)-2\Gamma\tau_c-2\gamma(1-\gamma)}{2\Gamma\tau_c+2\gamma(1-\gamma)}M_2^2,\end{eqnarray}
where in the last passage we have introduced the constant
$C=M_4/M_2^2$. Again one can wonder the ratio between energy
fluctuations and the square of granular temperature, obtaining:
\begin{equation}
\frac{\langle E^2(t)\rangle-\langle E(t)\rangle^2}{T_g(t)^2}=\frac{N}{2}\frac{2\Gamma \tau_c +\gamma(1-\gamma)[2+\gamma(\gamma+5)]}{2\Gamma \tau_c+\gamma(1-\gamma)[2-\gamma(1-\gamma)]}
\end{equation} 
which yields the value $N/2$ in the elastic case~\cite{fabii,visco}.

Switching to the reduced variable $c^2=v^2/(2M_2)$
we can look for an expression of the distribution function
in terms of Sonine polynomials:
\begin{equation}
f_1(c)\simeq\frac{1}{\sqrt{\pi}}e^{-c^2}[1+
s_{2} S_2(c^2)+..]
\label{fc}
\end{equation}
with  $s_2=-1+M_4/3 M_2^2$
and $S_2(c^2)=\frac{3}{8}-\frac{3}{2}c^2+\frac{1}{2}c^4$.
In practice, one approximates the
series (\ref{fc}) with a finite number of terms and since the leading
term is the Maxwellian, the closer the system to the elastic limit,
the less term suffice to describe the state. 
In the same spirit we assume  the following expansion for
the two-particle distribution function
expression
\bea
h_2(c_1,c_2)&=&-f_1(c_1)f_1(c_2)+\frac{1}{\pi}\exp{\left[-(c_1^2 +c_2^2)\right] }
[A_1 c_1 c_2\nonumber \\&+&A_2(\frac{1}{2}-c_1^2) (\frac{1}{2}-c_2^2)
+A_3 (c_1 c_2^3+ c_1^3 c_2)]
\label{fc2b}
\eea where the coefficients $A_i$ satisfy the relations: \bea A_1&=&
\frac{8Q_{11}}{M_2}-\frac{2Q_{13}}{M_2^2}\\ A_2&=&
1+\frac{Q_{22}}{M_2^2}\\ A_3&=&
-\frac{2Q_{11}}{M_2}+\frac{2Q_{13}}{3M_2^2} \eea  A
straightforward computation shows that $A_i \to 0$ (for $i=1,2,3$) in
the elastic limit $\alpha \to 1$ and in the Brownian limit $\Gamma
\tau_c \to \infty$, i.e. when the collision rate is so small that
grains thermalize with the external bath.

\section{Conclusion}
\label{Conclusion}

We have shown that if the number, $N$, of particles, experiencing inelastic
collisions described by the Inelastic Maxwell Model, is finite
it is possible to observe correlations of order $1/N$ among the velocities 
of different particles. Such correlations have been studied in two relevant
situations: the homogeneous cooling state and the steady state obtained by
applying a stochastic driving to the system.
In the first case we have obtained the velocity correlations by solving 
to order $1/N$ the
equations for the moments of the one and two-particles distribution
functions which show that the energy fluctuations decrease slower than the
squared energy.
In addition, we have studied the velocity pair correlation function 
in the scaling regime 
where the one-particle probability distribution is given by
$f_1(u,t)=2[\pi v_0(t)]^{-1}[1+(u/v_0(t))^2]^{-2}$.
 For small inelasticity 
we have obtained its explicit expression.
Interestingly, such a solution shows that the moment $Q_{22}$ of the velocity
pair distribution function diverges.
We may conjecture that such tails, which are the fingerprint 
of the Maxwell model, will persist in the many-particles 
correlation functions of higher order. These could be in principle
computed using the same methods discussed above, although the effort
required to carry out the program could be exceedingly heavy.

Finally, we have obtained, by the series expansion method,  the
pair distribution function when the system is subjected to a Langevin driving.
In this case the moments of the pair correlation are finite 
up to the fourth order and we believe that the higher moments will also
be finite.

As final remark we would like to comment that although the Maxwell model
is somehow artificial and does not describe any real granular material
it offers, as our paper illustrates, the possibilty of exploring
new aspects of non equilibrium statistical systems.   
\noindent {\it Acknowledgments.--} U.M.B.M. acknowledges the support of the 
Project COFIN-MIUR 2005, 2005027808.
\appendix
\section{Projection technique}
\label{appendix1}
Since we are not able to find a solution in the full space we resort
to an approximate method in the remaining Fourier space.
The method consists of projecting the term onto the subspace  
spanned by the function $\hat \xi_n$. 
For the sake of simplicity we define the scalar product between two functions
$f$ and $g$
\be
(f,g)=\int_{-\infty}^{+\infty} dq f(q) g(q) 
\ee  
and introduce an orthogonal basis, $R_n(q)$, of the form
\be
R_n(q)=\sum_{l=1}^n A_{nl}|q|^l\exp(-|q|)
\ee
The orthonormalization conditions for $q_1 q_2<0$ give the following relations:
\bea
R_0(|q|)&=& 2^{1/4}(\xi_0-\xi_1+\xi_2)\\
R_1(|q|)&=& 2^{1/4}(\xi_0-3 \xi_1+3 \xi_2)\\
R_2(|q|)&=& 2^{1/4}(\xi_0-5 \xi_1+7\xi_2)
\eea
in compact form:
\be
R_m(|q|)=\sum_n M_{mn}\xi_n
\label{orto}
\ee
We define, using the Heaviside function $\theta(x)$, 
$\tilde U(q_1,q_2)=[1-\theta(q_1 q_2)]U(q_1,q_2)$
 and expand with respect to the $R_n$'s
\be
\tilde U(q_1,q_2)=\sum_{a,b=0}^2 U_{ab} R_a(q_1) R_b(q_2)+K(q_1,q_2)
\label{a7}
\ee
where 
\be
U_{ab}= \int_{-\infty}^{+\infty}\int_{-\infty}^{+\infty} dq_1 dq_2 R_a(q_1) R_b(q_2) \tilde U(q_1,q_2)
\ee
and $K(q_1,q_2)$ represents the part of the function $\tilde U(q_1,q_2)$
orthogonal to the subspace spanned by the three functions above.

Inserting the ansatz 
$\chi_2(q_1,q_2)=\sum_{a,b}C_{ab}\xi_a(q_1)\xi_b(q_2)$
into \eqref{qre3}, 
neglecting the term $K(q_1,q_2)$ and using \eqref{orto}
we obtain
\be
[\Lambda_1 +\Lambda_2]\sum_{a,b}C_{ab}\xi_a(q_1)\xi_b(q_2)
=\sum_{m,n=0}^2 T_{mn}\xi_m(q_1)\xi_n(q_2)
\label{new2}
\ee
where $T_{mn}=\sum_{a,b=0}^2 U_{ab} M_{am}M_{bn}$.
Substituting the expansion
of $T_{mn}$ up to third order in $\zeta$  
we find the following  equation for the coefficients $C_{mn}$:
\bea
&\zeta&\Big [
C_{01}\xi_0(q_1)\xi_1(q_2)+C_{00}\xi_0(q_1)\xi_2(q_2)
+C_{10}\xi_1(q_1)\xi_0(q_2)
+2 C_{11}\xi_1(q_1)\xi_1(q_2)\nonumber\\
&+&(C_{10}+C_{12})\xi_1(q_1)\xi_2(q_2)
+C_{00}\xi_2(q_1)\xi_0(q_2)+
[C_{01}+C_{21}] \xi_2(q_1)\xi_1(q_2)\nonumber\\
&+&[C_{02}+C_{20}]\xi_2(q_1)\xi_2(q_2)\Big ]
=7\zeta^3\xi_0(q_1)\xi_0(q_2)\nonumber\\&-&(\frac{3}{2}\zeta^2+33\zeta^3)
(\xi_0(q_1)\xi_1(q_2)+\xi_1(q_1)\xi_0(q_2))
+(-2\zeta+21\zeta^2+135\zeta^3)\xi_1(q_1)\xi_1(q_2)\nonumber\\
&+&(-\zeta+\frac{3}{2}\zeta^2+45\zeta^3)(\xi_0(q_1)\xi_2(q_2)+
\xi_2(q_1)\xi_0(q_2))\nonumber\\
&+&(\zeta-32\zeta^2-161\zeta^3)
(\xi_2(q_1)\xi_1(q_2)+\xi_2(q_2)\xi_1(q_1))\nonumber\\
&+&(39\zeta^2+147\zeta^3)\xi_2(q_1)\xi_2(q_2)
\label{ex3}
\eea
whose solution is given by eqs. \eqref{c00}-\eqref{c02}

\section{Small terms}
In order to validate our assumption of neglecting the $1/N$ terms, we
have evaluated the terms on the r.h.s. of eq. (\ref{de1}) using the
relations given by eqs. \eqref{fmoments}-\eqref{qmoments}. 
We obtain that the $1/N$ terms are null up to third
order and the corresponding $k_1^4$ terms are explicitly
\bea
\hat f_1(\gamma k_1,\tau)\hat f_1((1-\gamma) k_1,\tau)-f_1( k_1,\tau) \nonumber\\\approx
...+\frac{1}{12}\Big\{\zeta^2\big[M_4(0)+3M_2^2(0)\big] e^{-a_4\tau} -2\zeta M_4(\tau)\Big\}k_1^4+...
\label{term2}
\eea
\bea
\frac{1}{N}\Big[\hat h _2(\gamma k_1,(1-\gamma)k_1,\tau) - \hat h _2(k_1,0,\tau)\Big] \nonumber\\ \approx
 \frac{1}{4N}\Big \{\zeta^2 \big[ Q_{22}(0)-M_4(0)\big] e^{-2a_2\tau} +2\zeta^2 M_4(\tau) \Big \}k_1^4+...
\label{termN2}
\eea
Comparing both coefficients we can state that our assumption 
is accurate at least up to fourth order.\\


\begin{thebibliography}{0}

\bibitem{introgas} P\"oschel T and Luding S (eds.), {\em Granular
Gases}, 2001 Lecture Notes in Physics, Vol. 564, Springer, Berlin

\bibitem{brey97} Brey J J, Dufty J W and Santos A, {\em Dissipative
dynamics for hard spheres}, 1997 {\em J. Stat. Phys} {\bf 87} 1051

\bibitem{noije} van Noije T P C and Ernst M H, {\em Velocity
Distributions in Homogeneously Cooling and Heated Granular Fluids},
1998 {\em Granular Matter} {\bf 1} 57

\bibitem{Baldassarri} Baldassarri A, Marconi U M B, and Puglisi A,
{\em Influence of correlations on the velocity statistics of scalar
granular gases}, 2002 {\em Europhys. Lett.} {\bf 58} 14 and {\em
Kinetic Models of Inelastic Gases}, 2002 {\em
Math. Mod. Meth. Appl. Sci.} {\bf 12} 965

\bibitem{Pugliomixturesundriven} Marconi U M B and Puglisi A, {\em
Mean-field model of free-cooling inelastic mixtures}, 2002 {\em
Phys. Rev. E} {\bf 65} 051305

\bibitem{BenNaim} Ben-Naim E and Krapivsky P L, {\em Multiscaling in
inelastic collisions}, 2000 {\em Phys. Rev. E} {\bf 61} R5

\bibitem{Santos} Santos A, {\em Transport coefficients of
d-dimensional inelastic Maxwell models}, 2002 {\em Physica A} {\bf
321} 442

\bibitem{Brito} Ernst M H and Brito R, {\em High-energy tails for
inelastic Maxwell models}, 2002 {\em Eurohys. Lett.} {\bf 58} 182

\bibitem{Goldhirsch} Goldhirsch I and Zanetti G, {\em Clustering
instability in dissipative gases}, 1993 {\em Phys. Rev. Lett.} {\bf
70} 1619

\bibitem{Sela} Sela N and Goldhirsch I, {\em Hydrodynamic equations
for rapid flows of smooth inelastic spheres, to Burnett order}, 1998
{\em J. Fluid Mech.} {\bf 361} 41

\bibitem{Andreabaldassarri} Baldassarri A, Marconi U M B, and Puglisi
A, {\em Cooling of a lattice granular fluid as an ordering process},
2002 {\em Phys. Rev. E} {\bf 65} 051301

\bibitem{vanNoije} van Noije T P C, Ernst M H, Brito R, Orza J A G,
{\em Mesoscopic Theory of Granular Fluids}, 1997 {\em
Phys. Rev. Lett.} {\bf 79} 411

\bibitem{Brey98} Brey J J, Moreno F and Ruiz-Montero M J, {\em Spatial
correlations in dilute granular flows: A kinetic model study}, 1998
{\em Phys.Fluids} {\bf 10} 2965

\bibitem{Soto} Soto R, Piasecki J and Mareschal M, {\em Precollisional
velocity correlations in a hard-disk fluid with dissipative
collisions}, 2001 {\em Phys. Rev. E} {\bf 64} 031306

\bibitem{Brey} Brey J J, Garc\'ia de Soria M I, Maynar P and
Ruiz-Montero M J, {\em Energy fluctuations in the homogeneous cooling
state of granular gases}, 2004 {\em Phys. Rev. E} {\bf 70} 011302

\bibitem{Bobylev}
Bobylev A V, {\em Exact solutions of the Boltzmann equation}, 1976 {\em Sov. Phys. Dokl.} {\bf 20} 820

\bibitem{Pagnani} Pagnani R, Marconi U M B and Puglisi A, {\em Driven
low density granular mixtures}, 2002 {\em Phys. Rev. E} {\bf 66}
051304

\bibitem{Andrea2002} Marconi U M B and Puglisi A, {\em Steady-state
properties of a mean-field model of driven inelastic mixtures}, 2002
{\em Phys. Rev. E} {\bf 66} 011301

\bibitem{Brito2002} Ernst M H and Brito R, {\em Driven inelastic
Maxwell models with high energy tails}, 2002 {\em Phys. Rev. E} {\bf
65} 040301(R)

\bibitem{Santos2003} Santos A and Ernst M H, {\em Exact steady-state
solution of the Boltzmann equation: A driven one-dimensional inelastic
Maxwell gas}, 2003 {\em Phys. Rev. E} {\bf 68} 011305

\bibitem{fabii}
Cecconi F, Diotallevi F, Marconi U M B and  
Puglisi A, {\em Fluid-like behavior of a one-dimensional granular gas}, 2004 {\em J. Chem. Phys} {\bf 120} 35

\bibitem{visco} Visco P, Puglisi A, Barrat A, van Wijland F and Trizac
E, {\em Energy fluctuations in vibrated and driven granular gases},
2006 {\em Eur. Phys. J B} {\bf 51} 377

\bibitem{brey2} Brey J J, Dufty J W, Ruiz-Montero M J, {\em Linearized
Boltzmann Equation and Hydrodynamics for Granular Gases}, (2004) in
{\em Granular Gas Dynamics - Lecture Notes in Physics}, Spinger
(Berlin)




\end{thebibliography}
\end{document}